\documentclass[a4paper,nofootinbib]{revtex4}

\usepackage{amsmath,amssymb}
\usepackage{epsfig}
\usepackage{graphicx}

\newcommand{\be}{\begin{equation}}
\newcommand{\ee}{\end{equation}}
\newcommand{\bi}{\begin{itemize}}
\newcommand{\ei}{\end{itemize}}
\newcommand{\ii}{\item}
\newcommand{\ba}{\begin{eqnarray}}
\newcommand{\ea}{\end{eqnarray}}
\newcommand{\gc}{\gamma_c}
\newcommand{\f}{\frac}

\def\nn{\nonumber \\}

\def\g{\gamma}
\def\d{\partial}

\def\abar{\bar\alpha}

\begin{document}

\bibliographystyle{h-physrev4}

\title{An alternative scaling solution for high-energy QCD saturation with running coupling}

\author{Guillaume~Beuf}
\email{guillaume.beuf@cea.fr}
\affiliation{Institut de physique th{\'e}orique, CEA/Saclay, 91191 
Gif-sur-Yvette cedex, France
\\URA 2306, unit{\'e} de recherche associ{\'e}e au CNRS}

\begin{abstract}
A new type of approximate scaling compatible with the Balitsky-Kovchegov equation with running coupling is found, which is different from the previously known running coupling geometric scaling. The corresponding asymptotic traveling wave solution is derived. Although featuring different scaling behaviors, the two solutions are complementary approximations of the same universal solution, and they become equivalent in the high energy limit. The new type of scaling is observed in the small-$x$ DIS data.
\end{abstract}

\maketitle


\section{Introduction}

Due to intense soft gluon radiation, hadronic processes at high energy involves strong color fields which have a nonlinear evolution, with in particular saturation effects. The mixing between quantum and nonlinear effects makes this high energy behavior difficult to derive from the QCD Lagrangian. However, in the last decade, many progresses have been made, in particular for the case of dense-dilute collisions, such as deep inelastic scattering (DIS) or proton-nucleus collisions. Two dual formulations of the same renormalization group evolution with rapidity for such collisions have been built, \emph{i.e.} Balitsky's hierarchy \cite{Balitsky:1995ub} and the JIMWLK equation \cite{Jalilian-Marian:1997jx,Jalilian-Marian:1997gr,Jalilian-Marian:1997dw,Iancu:2000hn,Iancu:2001ad,Ferreiro:2001qy}. These equations resum, at the leading logarithmic (LL) accuracy, the soft gluon emissions, taking into account nonlinear high-density effects at all orders for one of the two colliding particles. Neglecting correlations in the dense regime, those two formalisms reduce to the Balitsky-Kovchegov (BK) equation \cite{Balitsky:1995ub,Kovchegov:1999yj,Kovchegov:1999ua}. In fact, numerical simulations of the JIMWLK equation show no significant differences from solutions of the BK equation \cite{Rummukainen:2003ns}.  The BK equation gives the evolution with rapidity of the dipole-target amplitude, appearing in the dipole factorization of DIS, for example.
The \emph{geometric scaling} \cite{Stasto:2000er} property,
\ba
\sigma_{DIS}(Q^2,Y)&\simeq&\sigma\left(s_f\left(\log (Q^2/\Lambda_{QCD}^2),Y\right)\right)\\
\textrm{with} \quad s_f(L,Y)&\equiv& L- \abar v Y \, ,\label{FCGS}  
\ea
seen in the DIS data at high rapidity, can be explained by the scaling $N(L,Y)=N(s_f(L,Y))$ of the weighted Fourier transform \cite{Kovchegov:1999ua} of the dipole-target amplitude, which is approximately verified by the generic solutions of the BK and of the Balitsky-JIMWLK equations \cite{Iancu:2002tr}. We use the notation $\abar\equiv N_c\ \alpha_s/\pi$, and $L\equiv \log ({\bf k_t}^2/\Lambda_{QCD}^2)$, ${\bf k_t}$ being the Fourier conjugate of the parent dipole size ${\bf r}$.

In the three last years, a lot of efforts have been devoted to the improvement of the saturation equations, in mainly two directions.
First, the calculation of next-to-leading logarithmic (NLL) contributions has started \cite{Kovchegov:2006vj,Balitsky:2006wa,Balitsky:2007wg}, which are certainly required to achieve precision on the theory side. Moreover, at the LL accuracy, the QCD coupling remains fixed. Hence, the BK equation at LL accuracy is not expected to give good results over a wide energy range, due to the lack of asymptotic freedom. NLL contributions provides however indications about the best way to implement running coupling in the saturation equations \cite{Kovchegov:2006vj,Balitsky:2006wa}. Running coupling effects significantly modifies the whole behavior of the solutions \cite{Iancu:2002tr,Mueller:2002zm,Triantafyllopoulos:2002nz,Munier:2003sj}, whereas NLL contributions to the kernel or to the nonlinear term become irrelevant at high enough energy \cite{Beuf:2007cw}.

Second, the study of the inclusion of Pomeron loops in the saturation equations, and of the fluctuation effects induced by them, has been very active. However, it has been recently realized that in the case of dense-dilute processes, running coupling effects reduce the fluctuation effects in the high energy limit \cite{Dumitru:2007ew,Beuf:2007qa}, and even kill them completely at phenomenologically relevant rapidities \cite{Dumitru:2007ew}. Therefore we won't discuss further those effects in the present paper.

For the case of inclusive DIS, one has to calculate the forward part of the dipole-target amplitude. Assuming impact parameter independance \cite{Kovchegov:1999ua}, or a factorized momentum-transfert dependence \cite{Marquet:2005zf}, the BK equation for the forward dipole-target amplitude writes
\be
\d_Y N(L,Y)= \abar \chi(-\d_L) N(L,Y) - \abar N^2 \, . \label{BKFC}
\ee
In the previous formula, $\chi(\gamma)\equiv 2 \Psi(1)\!-\!\Psi(\gamma)\!-\!\Psi(1\!-\!\gamma)$ is the characteristic function of the BFKL kernel. Considering the parent dipole relative ${\bf k_t}^2$ as the relevant scale for the coupling, one obtains the forward BK equation with running coupling
\be
\d_Y N(L,Y)= \f{1}{b L} \chi(-\d_L) N(L,Y) - \f{1}{b L} N^2 \, , \label{BKRC}
\ee
where $b=(11N_c - 2N_f)/12N_c$.

Our main tool to study Eqs. (\ref{BKFC},\ref{BKRC}) is the traveling wave method \cite{bramson,vanS}, borrowed from nonlinear physics. The BK equation \eqref{BKFC} shares indeed the same qualitative structure with the FKPP equation \cite{fish,KPP}, for which that method has been developped, with an instability of the vacuum, a diffusion term, and a nonlinear damping. Thanks to this, those two equations have both traveling wave solutions \cite{Munier:2003vc,Munier:2003sj}, with the following property. For any initial condition with a steep enough tail (which is the case in QCD due to color transparency) the  traveling wave solution will evolve to an universal asymptotic solution, loosing memory of the initial condition. This universality property constrains not only the limit of the solution, but also the behavior of the solution at the first two subleading orders. This whole universal behavior is known \cite{Munier:2004xu} for the BK equation with fixed coupling \eqref{BKFC}. In the running coupling case, similar universal asymptotic traveling wave solutions appears, although the convergence towards the asymptotic behavior is less spectacular. The solution has the approximate scaling property $N(L,Y)=N(s_g(L,Y))$ \cite{Iancu:2002tr} with the variable
\ba
s_g(L,Y)&=&L-\sqrt{\f{v Y}{b}} \label{SatScaleOldSc}\, ,
\ea
for large $L$ and $s_g \ll\sqrt{Y}$, which becomes exact in the large $L$ and large $Y$ limit.
This corresponds to a {\it geometric-scaling} property
with
$\log Q^2_s \sim \sqrt{Y}$, instead of $\log Q^2_s \propto Y$ for the fixed coupling {\it geometric-scaling} \eqref{FCGS}. In the running coupling case, the universal asymptotic solution is  known including the first subleading order \cite{Mueller:2002zm,Munier:2003sj} only \footnote{However, some unpublished preliminary results concerning the second subleading order have been obtained by St\'ephane Munier.}.

From the theory point of view, the BK equation with running coupling is the most efficient tool to derive analytically the dipole-target amplitude. However, numerical simulations \cite{Albacete:2004gw,Albacete:2007yr} of the BK equation with running coupling have pointed out that the scaling function is significantly steeper near the saturation scale in the running coupling case than in the fixed coupling case, whereas the running coupling solution from \cite{Mueller:2002zm,Munier:2003sj} and the fixed coupling solution have similar shapes, with in particular the same coefficient $\gc\simeq 0.62$ playing the role of an anomalous dimension. That fact seem to rise some doubts about the relevance of the previously known asymptotic solution of \eqref{BKRC}: either this solution is not the same one as in the simulations, either strong corrections have been missed. Moreover, the running coupling geometric scaling \eqref{SatScaleOldSc} has been shown compatible with DIS data \cite{Gelis:2006bs}, but not significantly better than the fixed coupling geometric scaling \eqref{FCGS}. The too narrow $Q^2$ range of the available low-$x$ data can explain partly that observation, but maybe not completely, as fixed coupling and running coupling solutions differ significantly. The purpose of the present paper is then to reexamine the problem of finding analytical approximate solutions of the BK equation with running coupling \eqref{BKRC}, and understanding their validity range.

For sake of simplicity, let us give the main results of our study, whose
complete derivation is rather technical and is given in the text. We perform a
search for scaling variables compatible with the equation \eqref{BKRC}.
As a main  result we find that, contrary to the fixed coupling case
\eqref{BKFC},  no exact scaling can be obtained, but only
approximative ones. We find  two different versions of  asymptotic scaling valid
in the vicinity of the saturation scale and above, including the previously
known one \eqref{SatScaleOldSc} and a new one
\ba
s_n(L,Y)&=&\f{L}{2}-\f{v Y}{2 b L}\label{SatScaleNewSc}\, ,
\ea
for large $L$ and $s_n \ll L$, which is not, rigorously speaking a {\it
geometric-scaling} variable since it cannot be rewritten (by exponentiation) as ${\bf k_t}^2/Q^2_s(Y)$. In Eqs. (\ref{FCGS},\ref{SatScaleOldSc},\ref{SatScaleNewSc}), the index $f$ stands for \emph{fixed coupling}, the index $g$ for \emph{geometric} and the index $n$ for \emph{new}.
The asymptotic traveling wave solution associated with the scaling variable
\eqref{SatScaleNewSc} is derived using the traveling-wave method
\cite{bramson,vanS}. It writes
\ba
N(L,Y)&\propto& e^{- \gc \bar{s}_n  +{\cal O}(\log
L)}
\ \left[\textrm{Ai}\left(\xi_1+ \f{\bar{s}_n}{(D L)^{1/3}}\right)  + \dots \right]\label{solLEnewscI} \\
\textrm{with} \quad \bar{s}_n&=&\f{L}{2}-\f{v_c Y}{2 b L}- \f{3 \xi_1}{4} (D
L)^{1/3}\label{SatScaleNewScI}\, ,
\ea
for large $L$ and $1\ll \bar{s}_n \ll \sqrt{L}$. The neglected terms $\dots$ are of order ${\cal O}(L^{-1/3})$ when $\bar{s}_n={\cal O}\left(L^{1/3}\right)$.
The parameters $\gc$, $v_c$, and $D$ are defined in the text (see Eqs. (\ref{defgc},\ref{defvc},\ref{coeffD})), and $\textrm{Ai}$ is the Airy function. The two asymptotic traveling wave solutions, associated with the scaling variables $s_g$ and $s_n$ are two complementary approximations of the same exact solution. In particular, they both include the universal features of the exact solution, but differ significantly concerning some non universal features, such as the shape of the wave front near the saturation scale. Having two different approximate solutions allows a better control on the validity range of each, and can help to estimate a part of the errors.

The plan of our analysis is the following.
In section \ref{sec:scalings}, a search for scaling variables compatible with
the BK equation with running coupling \eqref{BKRC} is performed. In section
\ref{sec:UnivAsymptScalingSol}, the asymptotic traveling wave solution (\ref{solLEnewscI},\ref{SatScaleNewScI}) associated
with the scaling variable \eqref{SatScaleNewSc} is derived using the traveling
wave method \cite{bramson,vanS}. The approximate asymptotic traveling wave solutions associated with the scaling variables $s_g$ \eqref{SatScaleOldSc} and $s_n$ \eqref{SatScaleNewSc} are compared in section \ref{sec:compa}. The
phenomenological relevance of the new type of scaling $s_n$ is shown in
section \ref{sec:Pheno}. The last section contains the conclusion and an outlook.


\section{Scaling variables from saturation with running coupling}
\label{sec:scalings}


\subsection{Preliminaries: Scaling for a medium-dependent FKPP equation}

The FKPP equation has been useful \cite{Munier:2003vc} to understand the scaling properties of the BK equation at fixed coupling. Therefore, before considering the BK equation with running coupling \eqref{BKRC}, let us study the following modified FKPP equation, mimicking running coupling effects
\be
\d_t F(x,t)= \frac{1}{x} \left[ \d_x^2 F(x,t) + F(x,t) -F^2(x,t) \right] \, , \label{iFKPP}
\ee
where the operator $\d_x^2 + {\mathbb I}$ plays the same role as  the
kernel $\chi(-\d_L)$ in \eqref{BKRC}.
Note that this equation might be interesting also as a mean field description of a reaction diffusion process in
statistical physics when a particular medium implies a   {\it time} dependence
decreasing with the {\it distance} $x$.

$F$ is a scaling solution associated with the scaling variable $s(x,t)$ if it writes
\be
F(x,t)=F_{s}(s(x,t)) \label{AnsScaling}\, .
\ee
Inserting the condition \eqref{AnsScaling} in \eqref{iFKPP}, one gets
\be
x (\d_t s(x,t)) F_{s}'(s) = (\d_x^2 s(x,t)) F_{s}'(s)+(\d_x s(x,t))^2 F_{s}''(s) + F_{s}(s)-F_{s}^2(s) \, .
\ee
As $F_{s}(s)-F_{s}^2(s)$, $F_{s}'(s)$ and $F_{s}''(s)$ are scaling functions, \emph{i.e.} depend only on $s$, the coefficients of these functions must be scaling functions, in such a way that $F(x,t)=F_{s}(s)$ remains really an exact scaling solution of \eqref{iFKPP}.
All in all, finding an exact scaling variable $s$ associated with an exact scaling solution \eqref{AnsScaling} of \eqref{iFKPP} is equivalent to solving the equations
\ba
x \d_t s(x,t)&=&f_1(s) \label{eq1iFKPPbis}\\
\d_x s(x,t)&=&f_2(s) \, , \label{eq2iFKPPbis}
\ea
$f_1$ and $f_2$ being some arbitrary functions of $s$.

In order to have a smooth scaling variable $s(x,t)$, one has also to impose that
\be
\d_t \d_x s(x,t) = \d_x \d_t s(x,t) \, .
\ee
Eqs. (\ref{eq1iFKPPbis},\ref{eq2iFKPPbis}) give
\ba
\d_t \d_x s(x,t)&=&\f{1}{x} \ f_1(s) \ f_2'(s) \label{dd1mFKPP}\\
\d_x \d_t s(x,t)&=&\f{1}{x} \ f_1'(s) \ f_2(s) - \f{1}{x^2} \ f_1(s) \, . \label{dd2mFKPP}
\ea
One concludes that the inhomogeneous FKPP equation \eqref{iFKPP} has no nontrivial exact scaling solution.

Note that for the standard FKPP equation, the same approach would give the scaling conditions (\ref{eq1iFKPPbis},\ref{eq2iFKPPbis}) without the $x$ prefactor in \eqref{eq1iFKPPbis}. Hence, the two conditions would be compatible, and give the analog of the geometric scaling variable with $\log Q^2_s(Y) \propto Y$ and nothing else.\\

Large $x$ approximate scaling solutions are interesting.
For $x$ large, one can drop the last term in \eqref{dd2mFKPP} without assuming $f_1(s)\equiv 0$.
Matching the $1/x$ terms in \eqref{dd1mFKPP} and \eqref{dd2mFKPP} gives
\ba
\f{f_1'(s)}{f_1(s)}=\f{f_2'(s)}{f_2(s)} \, ,
\ea
then $f_1(s)$ and $f_2(s)$ are proportional, which allows us to write
\ba
x \ \d_t s(x,t)&=&-\f{v}{2} f_2(s) \label{eq1cmFKPP} \quad \textrm{with $v$ any real number}\\
\d_x s(x,t)&=&f_2(s) \, . \label{eq2cmFKPP}
\ea
$f_2$ reflects the invariance of the Ansatz \eqref{AnsScaling} under reparametrizations. One can choose a specific $f_2$ function without loss of generality. Choosing $f_2(s)\equiv 1$, one has
\ba
x \ \d_t s(x,t)&=&-\f{v}{2}  \label{eq1cmFKPPbis} \quad \textrm{with $v$ any real number}\\
\d_x s(x,t)&=&1 \, . \label{eq2cmFKPPbis}
\ea
As discussed before, the equation \eqref{iFKPP} has no exact scaling solutions. Then, the equations \eqref{eq1cmFKPPbis} and \eqref{eq2cmFKPPbis} cannot be solved simultaneously beyond a large $x$ approximation.\\

Consider the following alternative

i) {\it An exact solution for \eqref{eq2cmFKPPbis} and an approximative one for
\eqref{eq1cmFKPPbis}}: One gets from \eqref{eq2cmFKPPbis} $s(x,t)=x-\phi(t)$, then
\be
x\ \d_t s(x,t)= - x\ \phi'(t) = - \phi'(t)\ \phi(t)\ \left(1+\f{s(x,t)}{\phi(t)}\right)\, .
\ee
\noindent Hence, the equation \eqref{eq1cmFKPPbis} is approximately verified if
\be
\phi'(t)\ \phi(t)\ \left(1+\f{s(x,t)}{\phi(t)}\right) \simeq \f{v}{2} \label{eq1dmFKPP}\, .
\ee
That equation can be solved for $s\ll \phi(t)$, as
\be
\phi(t)=\sqrt{v(t-t_0)},
\ee
$t_0$ being an integration constant.
\noindent Thus, we find the scaling
\be
s(x,t)=x-\sqrt{v (t-t_0)} \label{geomscsqrtY}\, ,
\ee
\noindent which is valid for $x$ large and $s(x,t)\ll x$. The second condition means that $t-t_0={\cal O}(x^2)$, hence this scaling can be valid only if $x$ and $t$ are both large.  This solution corresponds, in the case of the BK equation with running coupling, to the solution with a modified geometric scaling involving the square root of the rapidity, as proposed in \cite{Iancu:2002tr,Mueller:2002zm,Triantafyllopoulos:2002nz,Munier:2003sj}.

ii) {\it An exact solution for \eqref{eq1cmFKPPbis} and an approximative one for
\eqref{eq2cmFKPPbis}}: One gets from \eqref{eq1cmFKPPbis}
\be
s(x,t)=\varphi(x)-\f{v t}{2 x} \, .
\ee
Then, the equation \eqref{eq2cmFKPPbis} rewrites
\be
1 \simeq \varphi'(x) + \f{v t}{2 x^2} = \varphi'(x) + \f{\varphi(x)}{x}-\f{s(x,t)}{x} \label{eq2dmFKPP}\, .
\ee
As previously, one finds an approximate solution for $x$ large and $s(x,t)\ll x$, which is
\be
\varphi(x)=\f{x}{2}-\f{v t_0}{2 x},
\ee
$t_0$ being an integration constant. Hence, for $x$ large and $s(x,t)\ll x$ we have the approximate scaling variable
\be
s(x,t)=\f{x}{2}-\f{v (t-t_0)}{2 x} \label{newscaling}\, .
\ee


\subsection{QCD case: the Balitsky-Kovchegov equation with running coupling}

Let us now consider the BK equation with running coupling \eqref{BKRC} with the same approach. In order to search for scaling solutions
\be
N(L,Y)=N_{s}(s(L,Y)) \label{AnsScalingBK}\, ,
\ee
it is convenient to write the kernel of \eqref{BKRC} as
\be
\chi(-\d_L)=\sum_{p=0}^\infty \f{1}{p!}\ \chi^{(p)}(\g)\ (-\gamma-\d_L)^p \quad \textrm{with} \quad \gamma \in ]0,1[ \, .
\ee
Thus, if $(-\gamma\!-\! \d_L)s(L,Y)$ is a scaling function, then $(-\gamma\!-\! \d_L)^p N_{s}(s(L,Y))$ is a scaling function, and then $\chi(-\d_L) N_{s}(s(L,Y))$ is a scaling function too. 
Hence, the constraints
\ba
bL \d_Y s(L,Y)&=&f_1(s) \label{eq1BKRC}\\
\d_L s(L,Y)&=&f_2(s) \label{eq2BKRC}
\ea
are sufficient conditions to have a scaling solution $N_{s}(s(L,Y))$ of the equation \eqref{BKRC}, and they are likely to be necessary too. As the conditions \eqref{eq1BKRC} and \eqref{eq2BKRC} are similar to \eqref{eq1iFKPPbis} and \eqref{eq2iFKPPbis}, the results found for the equation \eqref{iFKPP} apply for the Balitsky-Kovchegov equation \eqref{BKRC}.

We can thus give the following conclusions.
\bi
\ii On the contrary to the fixed coupling case, the BK equation with running coupling seems to have no exact scaling solutions. But in fact it is not really a problem, as the travelling wave method requires only a family of asymptotic scaling solutions. Moreover, an exact scaling would depend on the behavior of the coupling at transverse scales of the order of $\Lambda_{QCD}$ or smaller. As the BK equation has been derived in perturbation theory, it should not give such non perturbative information. Thus, even if exact scaling solutions of \eqref{BKRC} exist, beyond the large $L$ approximation, they would be unphysical.
\ii The BK equation with running coupling \eqref{BKRC} has solutions with large $L$ scaling similar to to the solutions (\ref{geomscsqrtY},\ref{newscaling}) of the modified FKPP equation \eqref{iFKPP}. The region near the front, where $Y={\cal O}(L^2)$, is compatible with two types of scaling behavior at large $L$, either the geometric scaling depending on the square root of the rapidity
\ba
s_g&=&L-\sqrt{\f{v (Y-Y_0)}{b}} \label{SatScaleOldScPar}\, ,
\ea
as derived in \cite{Iancu:2002tr,Mueller:2002zm,Triantafyllopoulos:2002nz,Munier:2003sj}, or a new non-geometric scaling
\ba
s_n&=&\f{L}{2}-\f{v (Y-Y_0)}{2 b L}\label{SatScaleNewScPar}\, .
\ea
\ei
In both cases, the parameter $Y_0$ reflects subasymptotic contributions. We will sometimes drop it in the following of the present theoretical study, but it is relevant for phenomenological studies of scaling \cite{Gelis:2006bs,BPRS}.


\section{Traveling wave solution with the new scaling variable} \label{sec:UnivAsymptScalingSol}

In this section we apply the traveling wave method \cite{bramson,vanS} to derive the asymptotic solution of the BK equation with running coupling based on the scaling \eqref{SatScaleNewScPar} at large $L$ and $Y={\cal O}(L^2)$. The most technical part of the calculation is given in the Appendix \ref{sec:appA}. The calculation is similar to the ones done in \cite{Munier:2003sj}, in particular for the running coupling solution based on the scaling \eqref{SatScaleOldScPar}. However, we use a slightly different presentation of the method, doing the calculation step by step, instead of taking an Ansatz. It is useful in order to clarify the validity range of the solution, and to understand the dynamics hidden in the Ansatz.\\

The first step is to rewrite the equation \eqref{BKRC} in term of the scaling variable $s_n$.
In order to find the solution with both the scaling and the subleading scaling violations, one has to keep a second variable. It is possible to keep the \emph{effective time} $t$, following \cite{Mueller:2002zm,Triantafyllopoulos:2002nz,Munier:2003sj}, except that now $t\equiv \abar(L) (Y\!-\!Y_0)= (Y\!-\!Y_0)/bL$. However, it seems more interesting to keep $L$. On the one hand, one can verify that, in that case, this choice will drive us to neglect
less terms than the choice of $t$, giving hope to obtain a more accurate solution. On the other hand, it allows us to have an easier control on the validity range of our result, as our first approximation is the validity of perturbation theory \emph{i.e.} $b L\gg 1$. Note that for all positive $L$ and $Y\geq Y_0$, $s_n(L,Y)\leq L/2$, and that $s_n = L/2$ corresponds to $Y=Y_0$. Finally, the equation \eqref{BKRC} rewrites in terms of $s_n$ and $L$
\be
-\f{v}{2} \ \d_{s_n} \bar{N}(s_n,L)=\chi\left(-\left(1\!-\!\f{s_n}{L}\right)\d_{s_n}-\d_L\right) \bar{N}(s_n,L) - \bar{N}^2(s_n,L) \label{BKRCsc}
\ee
where $\bar{N}(s_n(L,Y),L) \equiv N(L,Y)$.

In the traveling wave formalism for nonlinear equations, one distinguishes different phase-space regions depending on the dynamical mechanism at work. Let us examine two important such regions where it is possible to make analytical predictions on the form of the solution. These two regions are the ``front interior'' and the ``leading edge'', following the standard notations of Ref.\cite{vanS}.

\subsection{``Front interior'' of the traveling wave solution:}

As discussed in section \ref{sec:scalings}, the scaling in $s_n$ is valid only at large $L$. Let us first study the ``front interior'' of the solution, \emph{i.e.} the region defined by $s_n={\cal O}(1)$. The solution of \eqref{BKRCsc} here admits a large $L$ expansion in the front interior
\be
\bar{N}(s_n,L)=\bar{N}_0(s_n)+\f{1}{L}\ \bar{N}_1(s_n) + {\cal O}\left(\f{1}{L^2}\right) \, . \label{FrontIntExp}
\ee
Inserting \eqref{FrontIntExp} in equation \eqref{BKRCsc}, one finds that $\bar{N}_0(s_n)$ verifies
\be
-\f{v}{2} \ \d_{s_n} \bar{N}_0(s_n)=\chi\left(- \d_{s_n}\right) \bar{N}_0(s_n) - \bar{N}_0^2(s_n) \label{BKRCFI0} \, .
\ee
This equation can be solved numerically for any relevant value of $v$. Nevertheless we will need later an analytic expression for the tail of $\bar{N}_0(s_n)$ in the dilute regime, \emph{i.e.} the solution of
\be
-\f{v}{2} \ \d_{s_n} \bar{N}_0(s_n)\simeq\chi\left(- \d_{s_n}\right) \bar{N}_0(s_n) \label{BKRCFI0lin} \, .
\ee
Factorizing the leading exponential decay (with some unknown parameter $\g$), one writes
\be
\bar{N}_0(s_n)=e^{-\g s_n} \ f_{FI}(s_n) \, , \label{factFI}
\ee
$f_{FI}(s_n)$ behaving as a power for $s_n$ large.  The equation \eqref{BKRCFI0lin} then becomes
\be
\f{\g \ v}{2} \ f_{FI}(s_n) -\f{v}{2} \ \d_s f_{FI}(s_n) =\chi(\g)\ f_{FI}(s_n) - \chi'(\g)\ \d_{s_n} f_{FI}(s_n) + \sum_{p=2}^\infty \chi^{(p)}(\g) \f{(-1)^p}{p!} \ \d_{s_n}^p f_{FI}(s_n) \, . \label{BKRCFI0linbis}
\ee
One gets from \eqref{BKRCFI0linbis} the dispersion relation
\be
v=\f{2\ \chi(\g)}{\g} \label{reldisp} \, .
\ee
Then, either $\g$ is arbitrary and all the derivatives of $f_{FI}$ have to vanish, either the first derivative is the only non-vanishing one, which gives
\be
v=2\ \chi'(\g) \label{reldisp2} \,
\ee
and thus $\g=\gc$, defined by
\be
\chi(\gc)=\gc\ \chi'(\gc) \, . \label{defgc}
\ee
The velocity of the wave corresponding to $\gc$ is
\be
v_c=\f{2\ \chi(\gc)}{\gc} \label{defvc} \, .
\ee
Hence, the solutions of the equation \eqref{BKRCFI0lin} are
\begin{eqnarray}
\bar{N}_0(s_n)&=& A \ e^{-\g s_n} \qquad \textrm{with} \quad \g\neq\gc \label{UnifTranslFront} \\
\textrm{or} \qquad \bar{N}_0(s_n)&=& A\ (s_n+B) \ e^{-\gc s_n} \label{UnivFrontIntSol} \, ,
\end{eqnarray}
with $A$ positive and $B$ real. The solutions \eqref{UnifTranslFront} are valid large $L$ solutions of \eqref{BKRCsc} for $s_n\ll \sqrt{L}$. When $s_n={\cal O}(\sqrt{L})$ the terms with $f_{FI}'$ come at the same order as some term suppressed with $s_n/L$, and the dynamics is modified. The solutions \eqref{UnivFrontIntSol} are the large $L$ critical solutions of \eqref{BKRCsc}. They are valid for $s_n\ll L^{1/3}$. New dynamical effects occur for ${\cal O}(s_n)={\cal O}(L^{1/3})$, when the term with $f_{FI}''$ comes at the same order as a term suppressed with $s_n/L$. This region is called the ``leading edge'' \cite{vanS} and will be studied in the next subsection.

Eqs.(\ref{UnifTranslFront},\ref{UnivFrontIntSol}) give both valid solutions of the linear part of the BK equation with running coupling \eqref{BKRC}. However, if the initial condition is steep enough, the nonlinear damping term will select dynamically \cite{bramson,vanS} the critical solution \eqref{UnivFrontIntSol}, and then the contributions of the non-critical solutions \eqref{UnifTranslFront} will not survive asymptotically. In that case, the resulting solution is called a \emph{pulled front}. By contrast, \emph{pushed fronts} are solutions in which the generic partial waves of the type \eqref{UnifTranslFront} would survive, for exemple if the initial condition were not steep or if the phase velocity given by the dispersion relation would have no minimum.

Going back to the equations \eqref{BKRCsc} and \eqref{FrontIntExp}, one can in principle calculate $\bar{N}_1(s_n)$ and higher order terms recursively. But we will stop here in this work for two reasons. First, $\bar{N}_1(s_n)$ is not needed for the following calculations at the accuracy considered in this paper. Second, $\bar{N}_0(s_n)$ is free from NLL corrections (beyond the resummed corrections giving the running coupling considered here), whereas $\bar{N}_1(s_n)$ would depend on NLL contributions, $\bar{N}_2(s_n)$ on NNLL contributions, \emph{etc}.


\subsection{``Leading edge'' of the traveling wave solution:}

We study now the equation \eqref{BKRCsc} when $L$ and $s_n$ are both large. We
drop the nonlinear term, because $s_n$ large corresponds to the tail of the front.
Let us factorize the exponential decay of the solution as in \eqref{factFI}, but
for the whole solution
\be
\bar{N}(s_n,L)=e^{-\g s_n} \ f_{LE}(s_n,L) \, . \label{factLE}
\ee
Then, expanding the kernel around $\g$, the equation \eqref{BKRCsc} becomes
\be
\f{v \g}{2} \ f_{LE}(s_n,L) - \f{v}{2}\ \d_{s_n} f_{LE}(s_n,L) = \sum_{p=0}^\infty
\f{1}{p!} \chi^{(p)}(\g) \left[-\left(1-\f{s_n}{L}\right)\d_{s_n} -\d_L - \f{\g s_n}{ L}
\right]^p f_{LE}(s_n,L) \, . \label{BKRCscLE1}
\ee
Hence for $s_n\ll L$, the cancelation of the leading term on each side gives
again the dispersion relation \eqref{reldisp}, and it remains
\be
- \f{v}{2}\ \d_{s_n} f_{LE}(s_n,L) = \sum_{p=1}^\infty \f{1}{p!} \chi^{(p)}(\g)
\left[-\left(1-\f{s_n}{L}\right)\d_{s_n} -\d_L - \f{\g s_n}{ L} \right]^p f_{LE}(s_n,L) \,
.\label{BKRCscLE2}
\ee
As we are interested by the solution selected by the front formation mechanism,
which corresponds to \eqref{UnivFrontIntSol}, $\d_{s_n} f_{LE}$ has to be non-zero.
Then, for $f_{LE} \times s_n/L\ll \d_{s_n} f_{LE}$, which corresponds \emph{a priori}
to $s_n\ll \sqrt{L}$, the leading terms in the equation \eqref{BKRCscLE2} gives
\be
- \f{v}{2}\ \d_{s_n} f_{LE}(s_n,L)= - \chi'(\g) \ \d_{s_n} f_{LE}(s_n,L)
\ee
Hence, the relation \eqref{reldisp2} holds while $s_n\ll \sqrt{L}$, which fixes
$\g=\gc$ and $v=v_c\equiv 2\chi(\gc)/\gc$. Then, the equation \eqref{BKRCscLE2}
becomes
\be
0 = \chi'(\gc)  \left[\f{s_n}{L}\d_{s_n} -\d_L - \f{\gc s_n}{ L} \right]
f_{LE}(s_n,L)+\sum_{p=2}^\infty \f{1}{p!} \chi^{(p)}(\gc)
\left[-\left(1-\f{s_n}{L}\right)\d_{s_n} -\d_L - \f{\gc s_n}{ L} \right]^p f_{LE}(s_n,L)
\, .\label{BKRCscLE3}
\ee
For $s_n\ll \sqrt{L}$, the leading terms in \eqref{BKRCscLE3} gives
\be
\chi'(\gc)  \left[ \d_L + \f{\gc s_n}{ L} \right] f_{LE}(s_n,L) =  \f{1}{2}\
\chi''(\gc)\ \d_{s_n}^2 f_{LE}(s_n,L) \, .\label{BKRCscLE4}
\ee
Hence, non-trivial effects appears for $s_n={\cal O}(L^{1/3})$, and $f_{LE}$ is
allowed to have an $L$-dependance constrained by
\be
\d_L \ f_{LE} \leq {\cal O}(L^{-2/3} \ f_{LE}) \, .
\ee
Thus, the leading term in $f_{LE}$ can be factorized as a prefactor $\exp(-3\gc
\beta L^{1/3})$ times a scaling function depending only on $s_n/L^{1/3}$, where
$\beta$ is a constant which have to be determined. This prefactor amounts to
give a subleading correction to the traveling-wave scaling. We are thus led to
define a more accurate scaling variable
\be
\bar{s}_n \equiv s_n + 3 \beta L^{\f{1}{3}} =\f{L}{2}-\f{v_c Y}{2 b L} + 3 \beta
L^{\f{1}{3}} \label{scalingS1} \, .
\ee
The rather technical derivation of the leading edge solution is given in
Appendix 1.

Finally, the universal asymptotic traveling wave solution of the BK equation
with running coupling associated with the scaling \eqref{SatScaleNewScPar} writes, for large $L$ and $1\ll \bar{s}_n \ll \sqrt{L}$

\ba
N(L,Y)&\propto& e^{- \gc \bar{s}_n  +{\cal O}(\log
L)} \ \textrm{Ai}\left(\xi_1+ \f{\bar{s}_n}{(D L)^{1/3}}\right)  \label{solLEnewsc1} \\
\textrm{with} \quad \bar{s}_n&=&\f{L}{2}-\f{v_c Y}{2 b L}- \f{3 \xi_1}{4} (D
L)^{1/3}\label{SatScaleNewSc1}\, .
\ea


\section{Comparison between the scaling solutions}
\label{sec:compa}

Let us recall the known results concerning the scaling variable $s_g$ \eqref{SatScaleOldScPar}. The beginning of the leading edge expansion of the traveling wave solution associated to $s_g$ has been calculated in \cite{Mueller:2002zm,Munier:2003sj}, and writes, in our notations
\ba
N(L,Y)&\propto& e^{- \gc \bar{s}_g +{\cal O}(\log Y)} \ \left[\textrm{Ai}\left(\xi_1+ \f{\bar{s}_g}{\left(D \sqrt{\f{v_c\ Y}{b}}\right)^{1/3}}\right)  +  \dots \right]\label{solLEoldsc1} \\
\textrm{with} \quad \bar{s}_g&=&L-\sqrt{\f{v_c\ Y}{b}}- \f{3 \xi_1}{4} \left(D  \sqrt{\f{v_c\ Y}{b}}\right)^{1/3}\label{SatScaleOldSc1}\, .
\ea
It is valid for large $L$ and $1\ll \bar{s}_g \ll Y^{1/4}$. The neglected terms $\dots$ are of order ${\cal O}(Y^{-1/6})$ when $\bar{s}_g={\cal O}\left(Y^{1/6}\right)$.

This section is devoted to the comparison of that solution with the solution \eqref{solLEnewscI} derived in the previous section.

\subsection{The saturation scale}

The saturation scale is the typical scale at which occurs the transition between the dilute and the dense regime, \emph{i.e.} between the linear and the nonlinear regime. As the transition is smooth, that definition for the saturation scale is ambiguous. For simplicity, we assume that the magnitude
of the solution \eqref{solLEnewsc1} is dominated by the exponential factor. It leads to define the logarithmic saturation scale $L_s(Y)$ by the implicit relation
\be
 s_1\left(L_s(Y),Y\right)-{\cal O}(\log L)\equiv  \zeta_{sat}\label{defSatScale}\, ,
\ee
with $\zeta_{sat}$ some parameter of order one. Then, the usual saturation scale $Q_s(Y)$ is defined by
\be
L_s(Y)\equiv\log\left(\f{Q_s^2(Y)}{\Lambda_{QCD}^2}\right)\label{LogSatScale}\, .
\ee

Equivalently, it is useful to consider the reciprocal function $Y_s(L)$ of $L_s(Y)$, \emph{i.e.} the saturation rapidity at a given transverse scale.  
Using our result \eqref{SatScaleNewSc}, one gets 
\be
Y_s(L)= \f{2 b L}{v_c}\left[\f{L}{2}-\f{3 \xi_1}{4}(D L)^{\f{1}{3}}+{\cal O}(\log L)- \zeta_{sat}\right] +Y_0 \, . \label{TrueSatRapNewScaling}
\ee
Note that the $\zeta_{sat}$ dependance, parametrizing the ambiguity in the definition of the saturation scale, is subleading compared to the universal terms, even the unknown one ${\cal O}(\log L)$. The subasymptotic $Y_0$ parameter is even less relevant. Then we just drop it in the following.
Inverting approximately the expansion \eqref{TrueSatRapNewScaling}, one finds
\be
L_s^2(Y)=\f{v_c Y}{b}+\f{3 \xi_1}{2} D^{\f{1}{3}} \left(\f{v_c Y}{b}\right)^{\f{2}{3}} + \left(\f{v_c Y}{b}\right)^{\f{1}{2}} \left[{\cal O}(\log Y) + 2 \zeta_{sat}\right] \, .
\ee
Finally, the square root of that expression gives
\be
L_s(Y)=\sqrt{\f{v_c Y}{b}}+\f{3 \xi_1}{4} \left(D \sqrt{\f{v_c Y}{b}}\right)^{\f{1}{3}} + {\cal O}(\log Y) +  \zeta_{sat}\label{logQsRCu1} \, ,
\ee
which is exactly the expression for the logarithmic saturation scale given by the other running coupling solution \eqref{SatScaleOldSc1}, neglecting the $Y_0$. Hence, although the two approximate asymptotic solutions of the BK equation with running coupling are different, their prediction for the saturation scale agree asymptotically. It is a first evidence that those two approximate solutions are approximations of the same exact solution. They agree concerning the first subasymptotic correction to the saturation scale, and they would presumably agree for the second one, which is still unknown in both cases. Indeed, these first corrections are constrained by the universality property of the asymptotic traveling waves solutions. For that reason, the two approximate solutions allow both to extract the leading terms of the exact logarithmic saturation scale at running coupling.\\

\begin{figure}
\begin{center}
\begin{tabular}{cc}
\includegraphics[width=9cm]{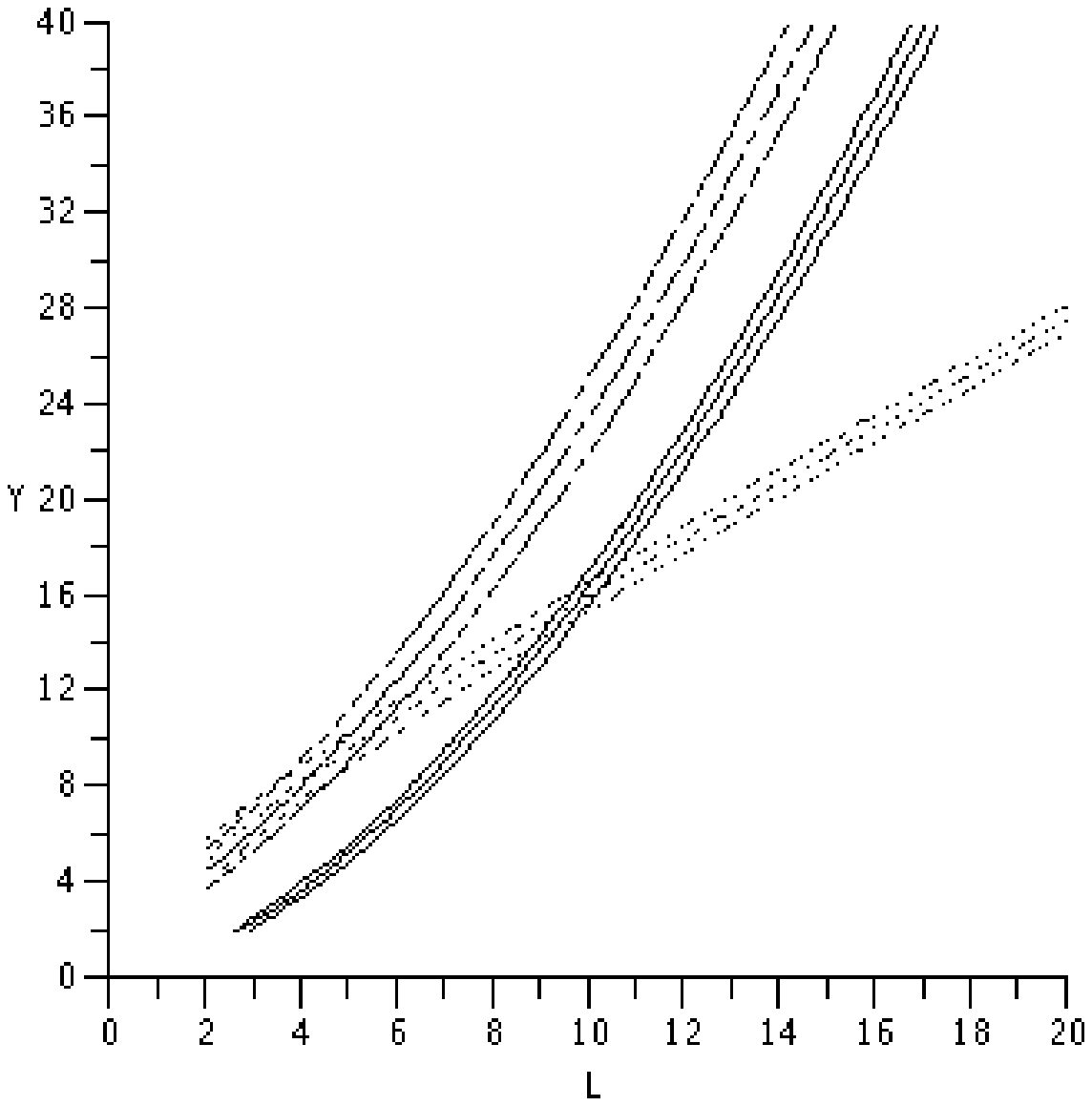} & \includegraphics[width=9cm]{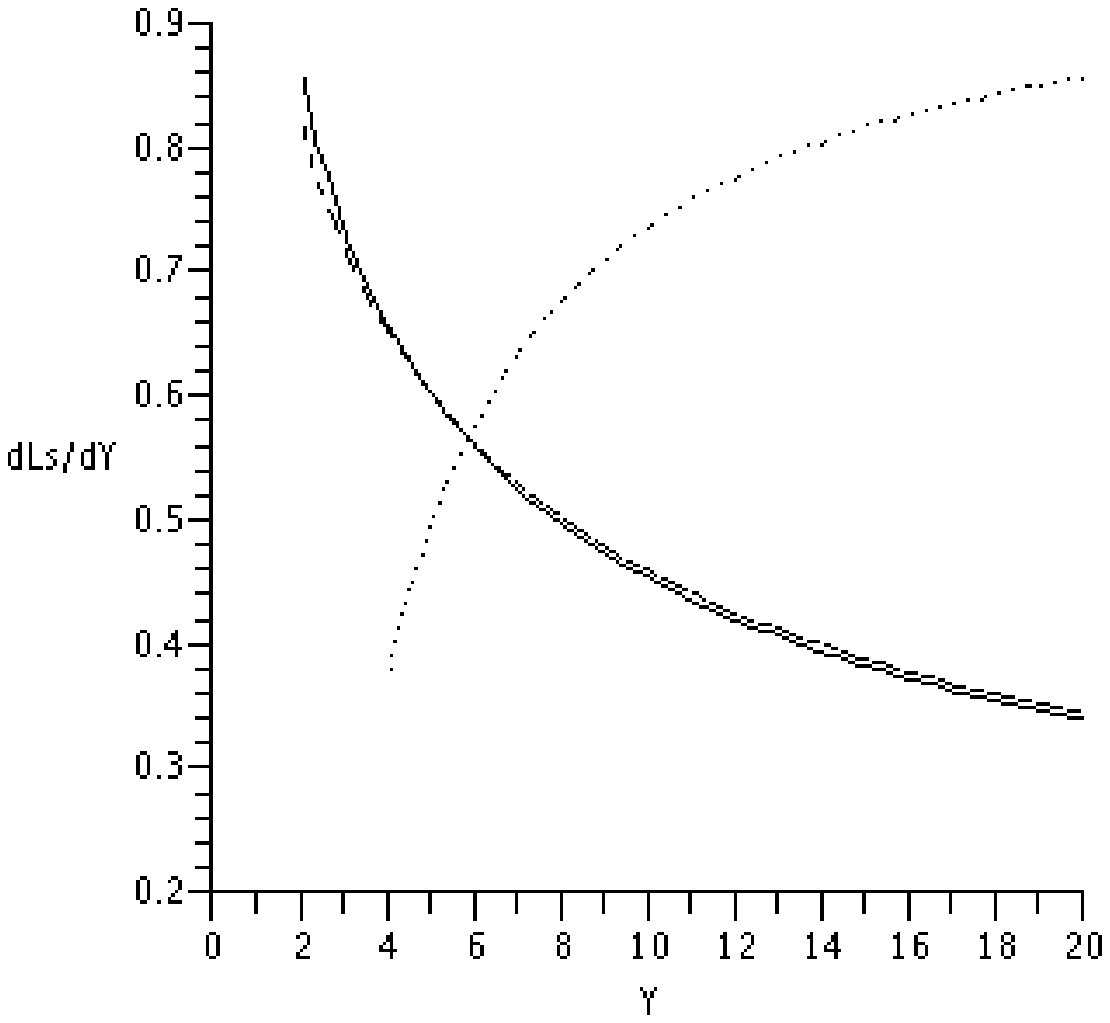}
\end{tabular}
\caption{\label{fig:SatScale} \emph{Saturation scale.} Solid lines: new running coupling scaling solution \eqref{TrueSatRapNewScaling}; Dashed lines: running coupling geometric scaling solution \eqref{logQsRCu1}; Dotted lines: fixed coupling geometric scaling solution. \emph{Left}: Saturation curve in the (L,Y) plane. For each scaling solution, the three curves corresponds (from the left to the right) to the values $0.5$, $1$ and $1.5$ for the parameter $\zeta_{sat}$ (see equation \eqref{defSatScale}).
\emph{Right}: Derivative of the logarithmic saturation scale $\textrm{d} L_s / \textrm{d} Y$ (see equation \eqref{LogSatScale}) as a function of the rapidity $Y$.}
\end{center}
\end{figure}

The behavior of the saturation scale determined from the two running coupling scaling variables $\bar{s}_g$ and $\bar{s}_n$ and from the corresponding fixed coupling scaling variable are compared in FIG. \ref{fig:SatScale}. The saturation line and the derivative of $L_s(Y)$ (see equation \eqref{LogSatScale}) are plotted using the expressions \eqref{logQsRCu1}, \eqref{TrueSatRapNewScaling}, dropping the ${\cal O}(\log Y)$ or ${\cal O}(\log L)$ terms, and taking $Y_0=0$. For each solution, the saturation line is plotted for three different values of the parameter $\zeta_{sat}$. As expected \cite{Iancu:2002tr}, at large $Y$ the saturation line is parabolic for running coupling and linear for fixed coupling. The two running coupling solutions agrees concerning the shape, but not concerning the normalization of $L_s(Y)$. It comes partly from the dropped terms, and partly from the ambiguity in the definition of the saturation scale. The parameter $\zeta_{sat}$ affects indeed more the the normalization of $L_s(Y)$ than its shape. The derivative of $L_s(Y)$ corresponds in the saturation models (\emph{e.g.} \cite{GolecBiernat:1998js} and many following models) to the parameter usually called $\lambda$. One remarks that the fixed coupling solution is not consistent with the fitted value $\lambda \simeq 0.3$, $\lambda$ being stable in the rapidity range covered by HERA. By contrast, such low and stable value can be provided in the running coupling case, as noticed in \cite{Triantafyllopoulos:2002nz}, but seems to require a large rapidity shift $Y_0$, which may be considered as a signal of important NLL corrections. The shape of the saturation scale is indeed less sensitive than the additionnal $Y_0$ parameter to higher orders.


\subsection{The shape of the front}

To compare the solutions \eqref{solLEoldsc1} and \eqref{solLEnewsc1} further, let us now consider the shape of $N(L,Y)$ obtained in both cases.
\begin{figure}
\begin{center}
\begin{tabular}{cc}
\includegraphics[width=9cm]{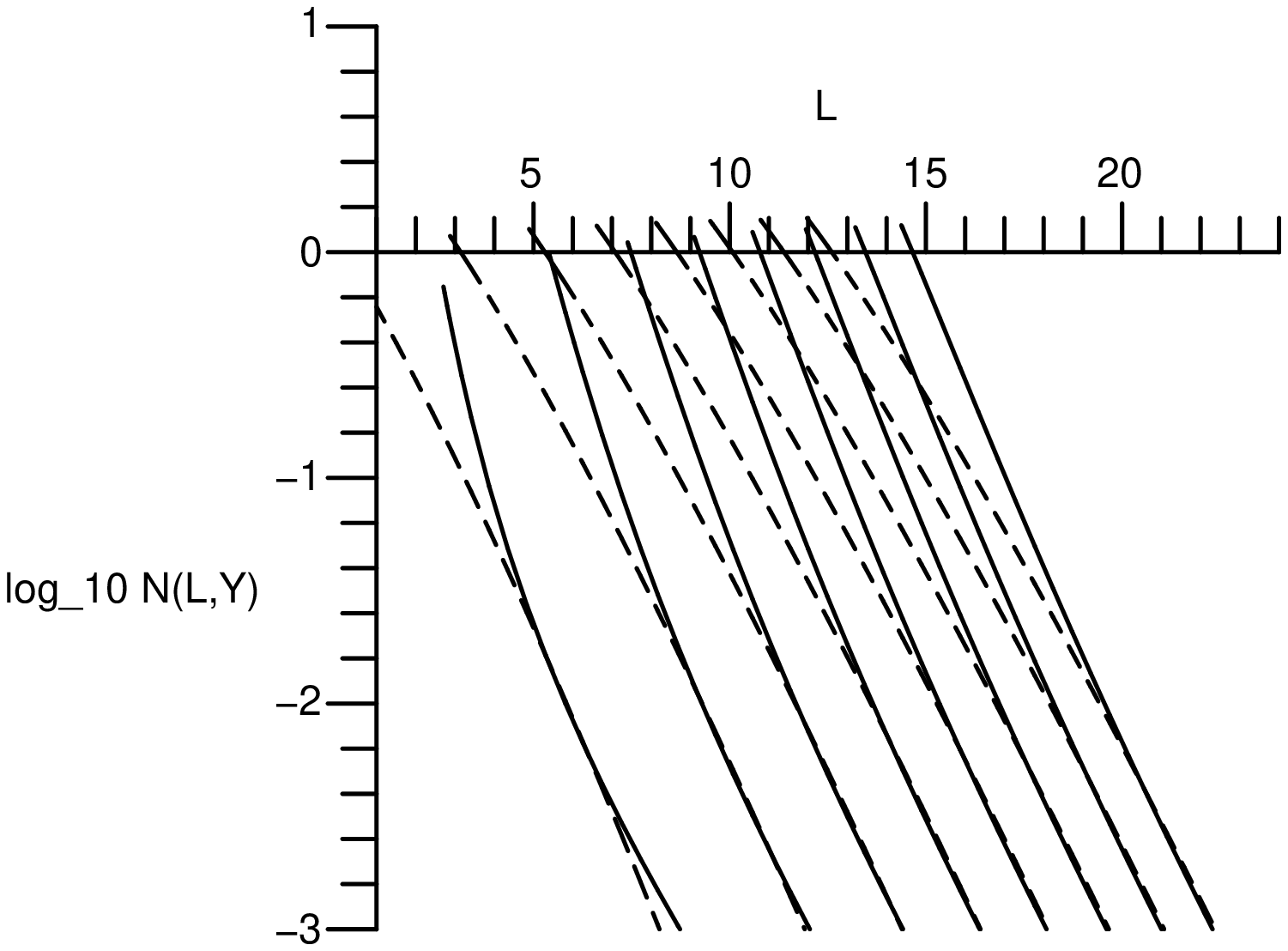} & \includegraphics[width=9cm]{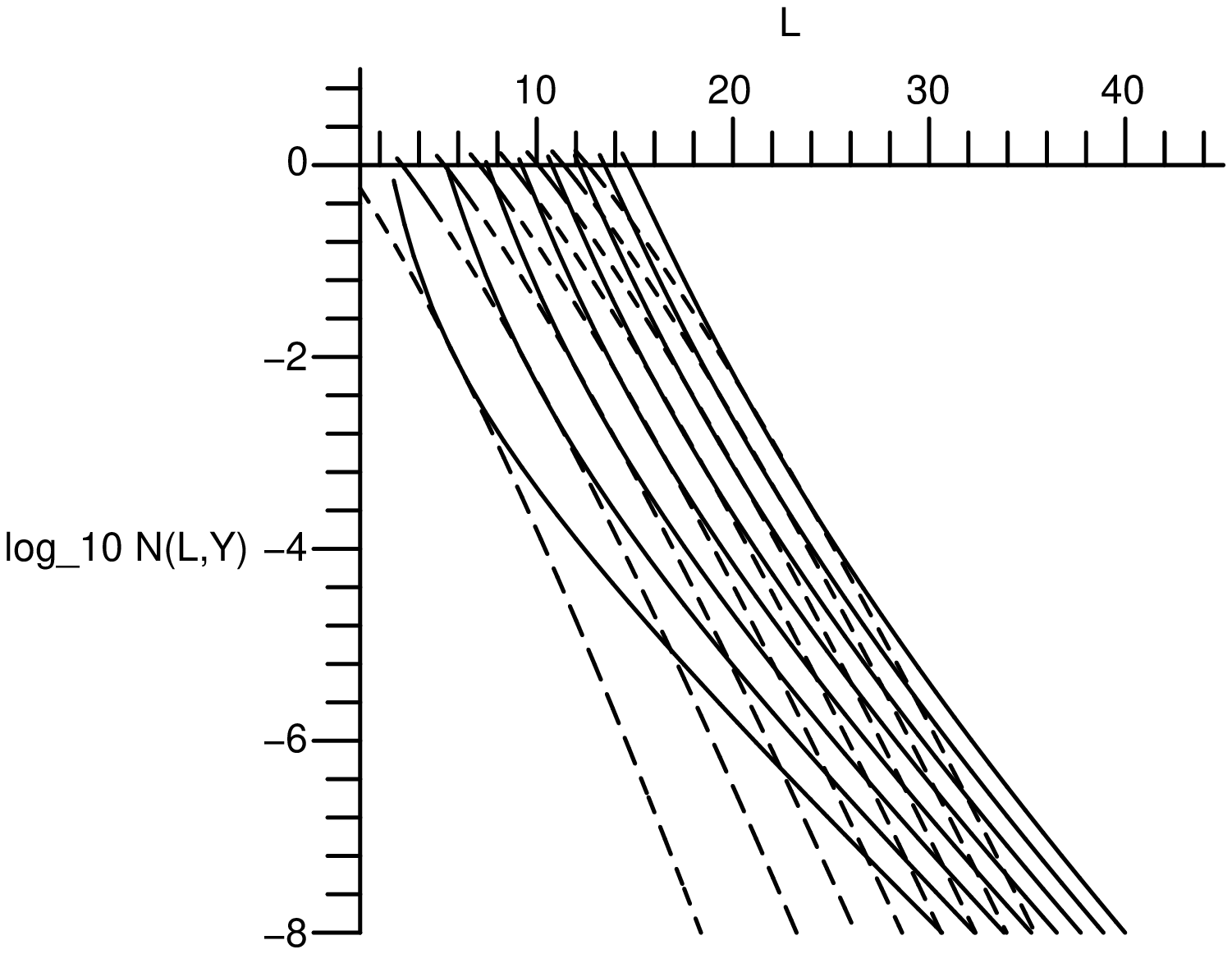}\\
\end{tabular}
\caption{\label{fig:RCshape} $N(L,Y)$ as a function of $L$ for the $s_g$ solution \eqref{solLEoldsc1} in dashed line, and for the $s_n$ solution \eqref{solLEnewsc1} in solid line.  From the left to the right the different curves corresponds to $Y=2$ to $Y=30$ by step of $4$. The vertical axis is in logarithmic (base 10) scale.  The two plots just differ by the range they cover.}
\end{center}
\end{figure}
They are plotted on FIG. \ref{fig:RCshape}, down to the vicinity of the saturation scale, but not beyond, as we have explicit solutions only in the linear regime. The values chosen for the parameters are $A=1$, $B=4$ and $Y_0=0$. The left hand plot corresponds to a range more relevant for phenomenology, but the right hand plot is interesting concerning the mathematical meaning of these two approximate solutions. They are both derived with the traveling wave method, as the beginning of an expansion in the leading edge region. As the two solutions coincide over a part of the linear region, it is obvious that they correspond to \emph{different expansions} of the \emph{same} exact solution of \eqref{BKRC}, \emph{i.e.} the exact pulled front solution selected dynamically. The range over which the two solutions agree is growing with $Y$ and drifting towards the tail of $N$, as expected for the leading edge region. Hence, the two approximate solutions \eqref{solLEoldsc1} and \eqref{solLEnewsc1} are in agreement concerning all the features constrained by the universality property of the pulled front solutions, \emph{i.e.} the evolution of the saturation scale and the shape of the front in the leading edge region. They give presumably an accurate description of these features, at least in the large $L$ limit (at $\bar{s}_g$ or $\bar{s}_n$ fixed). However they differ concerning other features, such as the shape of the front around the saturation scale (\emph{i.e.} in the front interior) or in the tail forward the leading edge. For exemple, asymptotic freedom effects are stronger for the $s_n$ solution
(\ref{solLEnewsc1},\ref{SatScaleNewSc1}) than for the $s_g$ solution
(\ref{solLEoldsc1},\ref{SatScaleOldSc1}), as the tail of the front evolves slower than the bulk in the former one. The left hand plot shows that near the saturation scale, the front given by the $s_n$ solution is significantly steeper than the one given by the $s_g$ solution. One can study the shape of the front more precisely by calculating an effective anomalous dimension defined as
\be
\gamma_{eff}(L,Y) \equiv - \d_L \log N(L,Y) \label{gammaeff}\, .
\ee
\begin{figure}
\begin{center}
\begin{tabular}{cc}
\includegraphics[width=9cm]{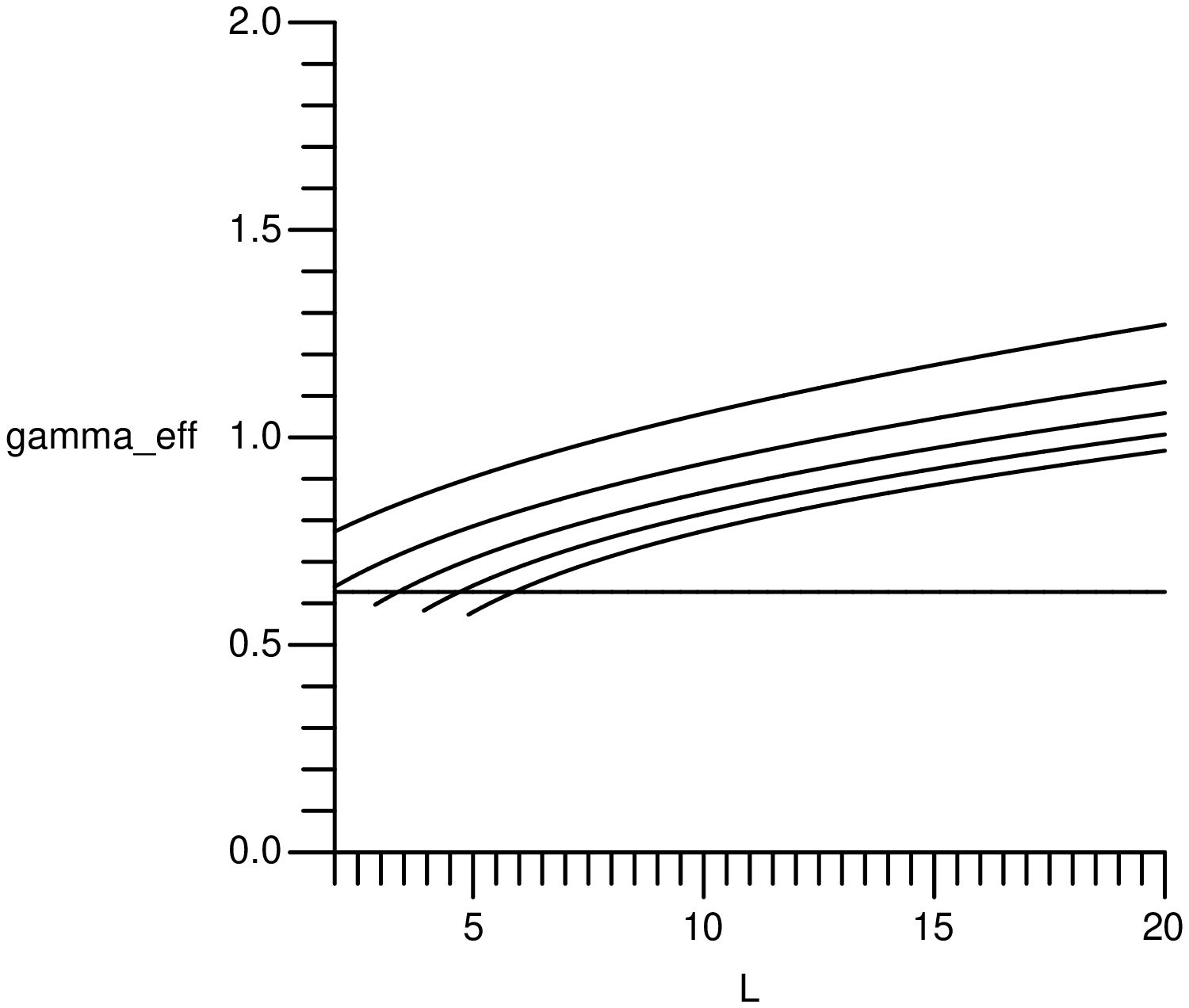} & \includegraphics[width=9cm]{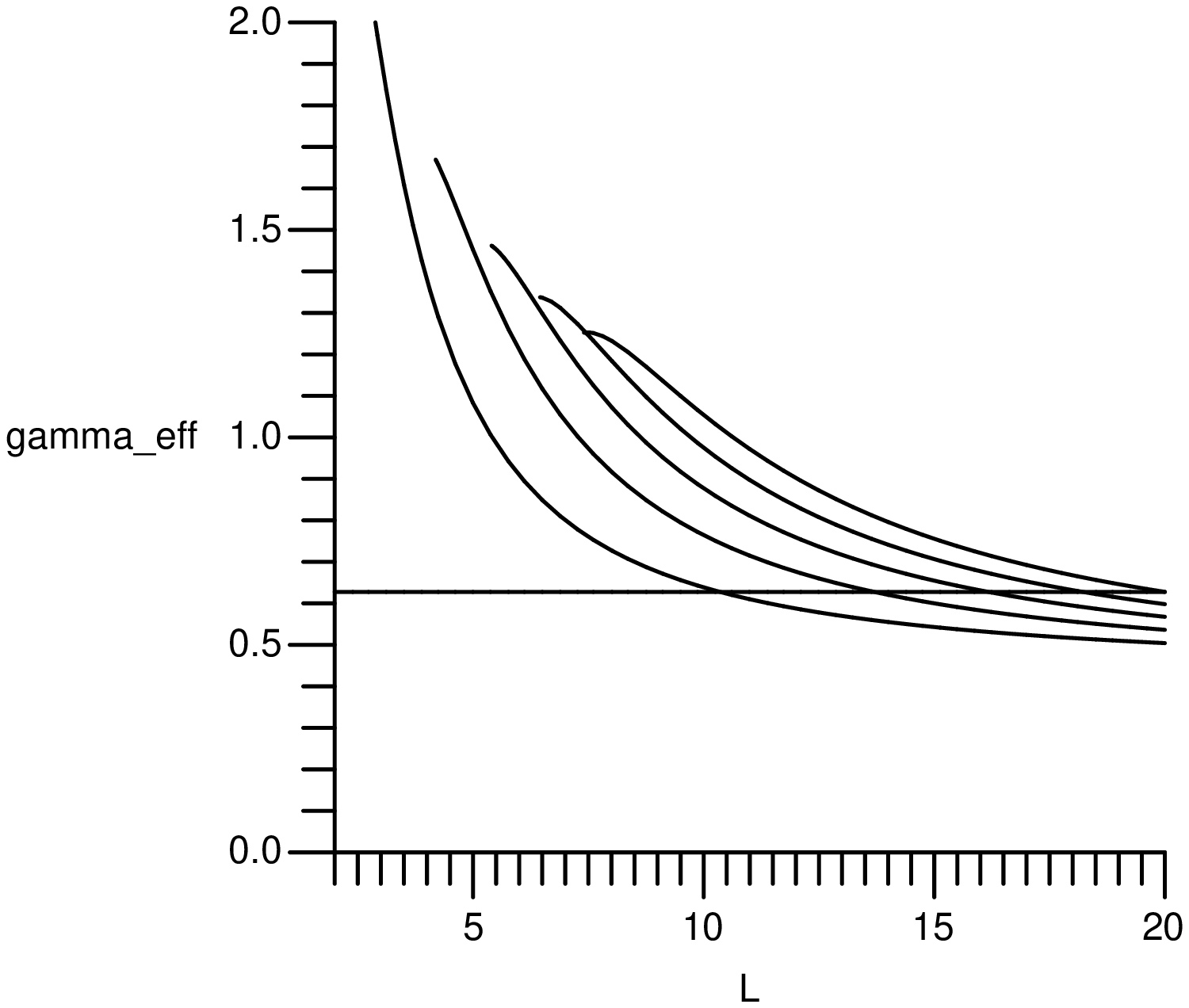}\\
\end{tabular}
\caption{\label{fig:gaeff} Effective anomalous dimension $\gamma_{eff}(L,Y)$ (as defined by equation \eqref{gammaeff}) as a function of $L$, calculated for the solutions $s_g$ solution (left panel) and $s_n$ solution (right panel). From the left to the right the different curves corresponds in both cases to $Y=2$ to $Y=10$ by step of $2$. The horizontal line stands for the value of the critical saturation exponent $\gc \simeq 0.63$.}
\end{center}
\end{figure}
The behavior of $\gamma_{eff}(L,Y)$ for the two running coupling approximate solutions is compared on FIG. \ref{fig:gaeff}. They give very different results. Moreover, both differ from the naive expectation $\gamma_{eff}(L,Y) \simeq \gc$. The factor containing the Airy fonction indeed contributes to the shape of the front in the leading edge, and the nontrivial structure of the scaling variable $s_n$ gives also a contribution to $\gamma_{eff}(L,Y)$ even in the front interior. Therefore, the theoretical solutions \eqref{solLEoldsc1} and \eqref{solLEnewsc1} (and the fixed coupling one) involve the exponent $\gc$, but it \emph{does not} mean that the anomalous dimension of these theoretical solutions is indeed $\gc$, especially at finite $L$ and $Y$.
\begin{figure}\label{fig:FrontShapesFCRC}
\begin{center}
\begin{tabular}{ccc}
\includegraphics[width=9cm]{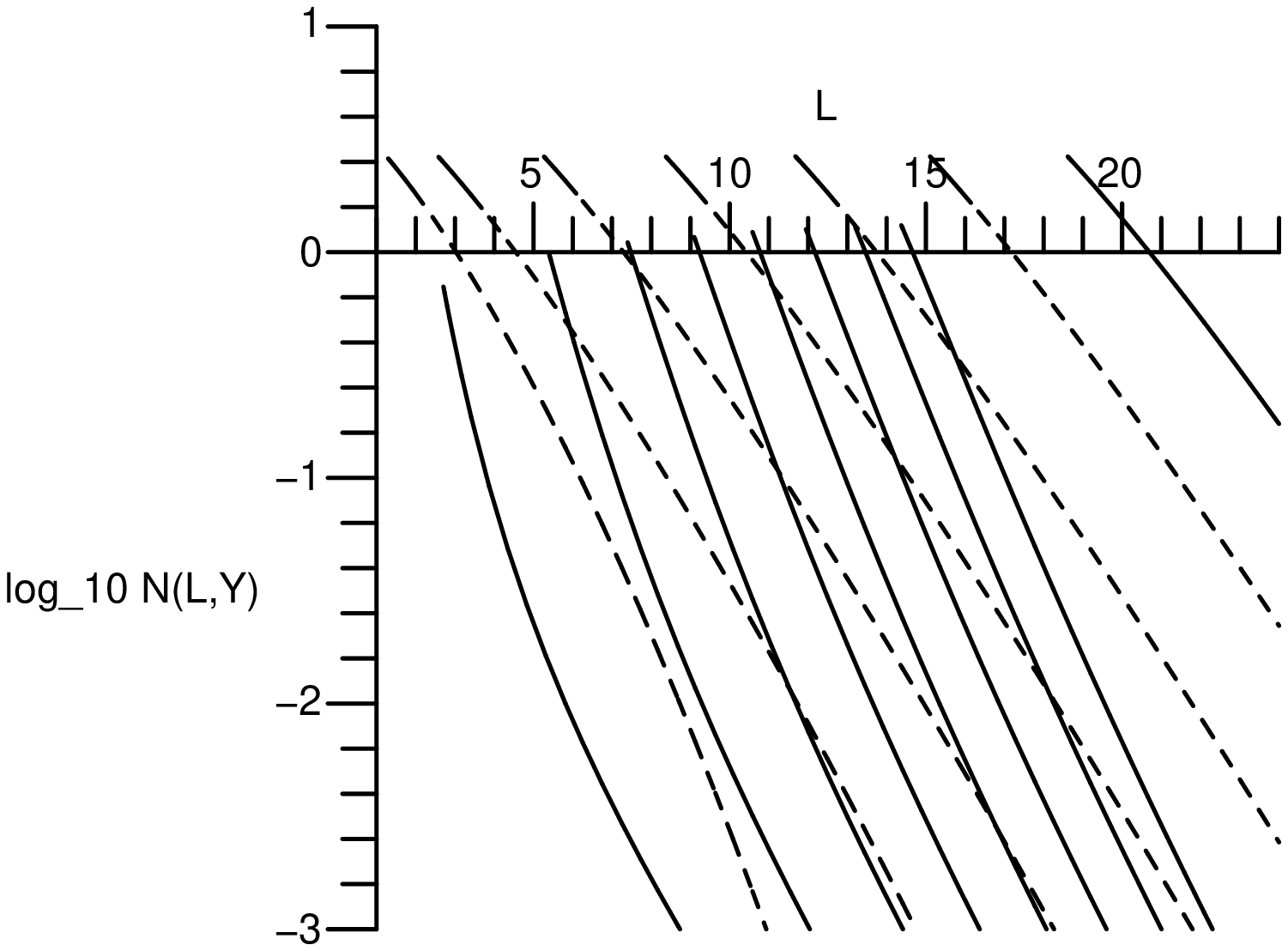} & & \includegraphics[width=9cm]{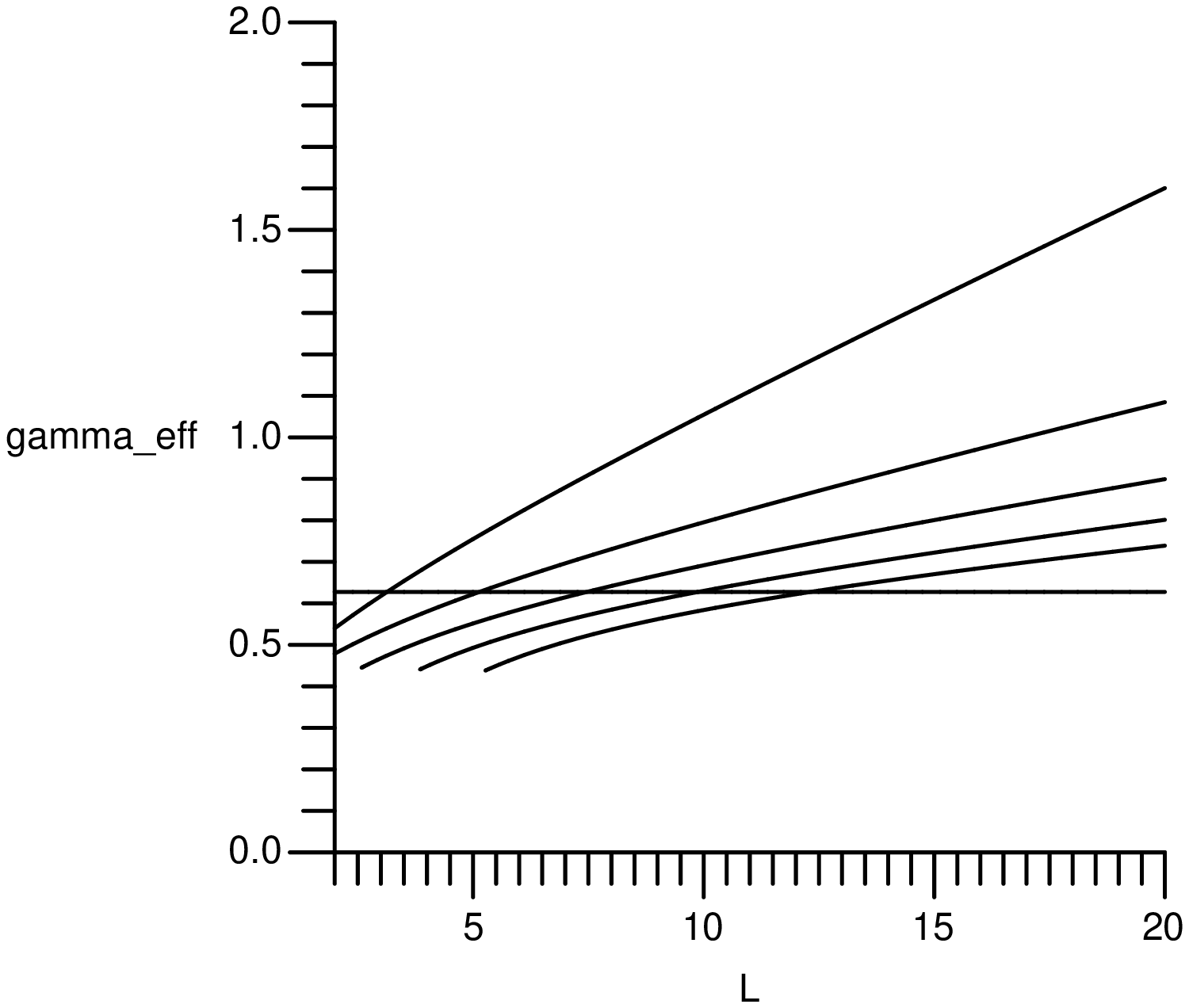}\\
\end{tabular}
\caption{\label{fig:FCshape} \emph{Left}: $N(L,Y)$ as a function of $L$ for the fixed coupling solution \cite{Mueller:2002zm,Munier:2003sj} in dashed line, and for the $s_n$ solution \eqref{solLEnewsc1} in solid line.  From the left to the right the different curves corresponds to $Y=2$ to $Y=30$ (or $Y=26$ for fixed coupling) by step of $4$. The vertical axis is in logarithmic (base 10) scale.
\emph{Right}: Effective anomalous dimension $\gamma_{eff}(L,Y)$ (as defined by equation \eqref{gammaeff}) as a function of $L$, calculated for the fixed solution. From the left to the right the different curves corresponds in both cases to $Y=2$ to $Y=10$ by step of $2$. The horizontal line stands for the value of the critical saturation exponent $\gc \simeq 0.63$.}
\end{center}
\end{figure}
For completeness, it is useful to recall the behavior of the universal travelling wave solution of the fixed coupling BK equation \eqref{BKFC}. This solution and the corresponding $\gamma_{eff}(L,Y)$ are plotted on FIG. \ref{fig:FCshape}. Its front has a shape very different from the one of the $s_n$ solution, but rather similar to the one of the $s_g$ solution.

These observations allow to address a puzzle raised by the numerical simulations of the BK equation with running coupling \cite{Albacete:2004gw,Albacete:2007yr}. It was noticed that numerical simulations leads to a wave front much steeper at running coupling than at fixed coupling, whereas one expect a similar wave front from the theoretical fixed coupling solution and $s_g$ running coupling solution \eqref{solLEoldsc1}. We have shown that the latter is only an approximate solution of \eqref{BKRC}, and that \eqref{solLEnewsc1} is another one, within a complementary approximation. Hence the exact pulled front solution of \eqref{BKRC} lies probably between these two approximate solutions, and can then have a stronger $\gamma_{eff}(L,Y)$ than the fixed coupling solution in the front interior.


\section{Phenomenology}
\label{sec:Pheno}

In this section, we discuss the phenomenological validity of the new scaling, using the results of Ref.\cite{BPRS}. The quality factor method \cite{Gelis:2006bs} allows to test and quantify scaling behaviors of experimental data. For a given data set, the higher is the quality factor value, the better is the scaling behavior. This method is used in \cite{BPRS} to compare scaling behaviors at the cross section level. Indeed, geometric scaling at the cross section level comes from geometric scaling properties of the dipole-target amplitude through dipole factorization and the replacement $k_t^2 \mapsto Q^2$ in the scaling variable. Note that such property is only approximate for the new scaling variable $s_n$.

Starting from the theoretical scaling variables $s_f$ \eqref{FCGS}, $s_g$ \eqref{SatScaleOldScPar} and $s_n$ \eqref{SatScaleNewScPar}, we shall use the following versions at the cross section level
\ba
\tau_f&=&\log \left(\f{Q^2}{\Lambda^2}\right)-\lambda\ Y\, , \label{FCpheno}\\
\tau_g&=&\log \left(\f{Q^2}{\Lambda^2}\right)-\lambda\ \sqrt{Y-Y_0}\, ,\label{RC1pheno}\\
\textrm{and} \quad \tau_n&=&\log \left(\f{Q^2}{\Lambda^2}\right)-\lambda\ \f{Y-Y_0}{\log \left(\f{Q^2}{\Lambda^2}\right)}\, . \label{RC2pheno}
\ea
$\lambda$ is taken as a free parameter in each case. The rapidity shift $Y_0$ is either fixed to zero or treated as a free parameter. Note that such shift cannot affect the quality of the scaling with the $\tau_f$ variable. The value of $\Lambda$ has an influence on the scaling quality only in the $\tau_n$ scaling case. $\Lambda$ is either fixed to $\Lambda_{QCD}=0.2$ GeV, or treated as a free parameter. 

Using the quality factor method on the inclusive DIS cross section data from H1, ZEUS, E665 and NMC experiments (see \cite{BPRS} for a complete discussion of data selection and references) in the range $3$ GeV$<Q^2<150$ GeV and $x<10^{-2}$, we have found the following results \cite{BPRS}. The optimal quality factor from one scaling variable to another, when $\lambda$ is the only free parameter, do not differ more than $4\%$. The data thus scale equally well with the three variables. 
\begin{figure}
\begin{center}
\begin{tabular}{cc}
\includegraphics[width=9cm]{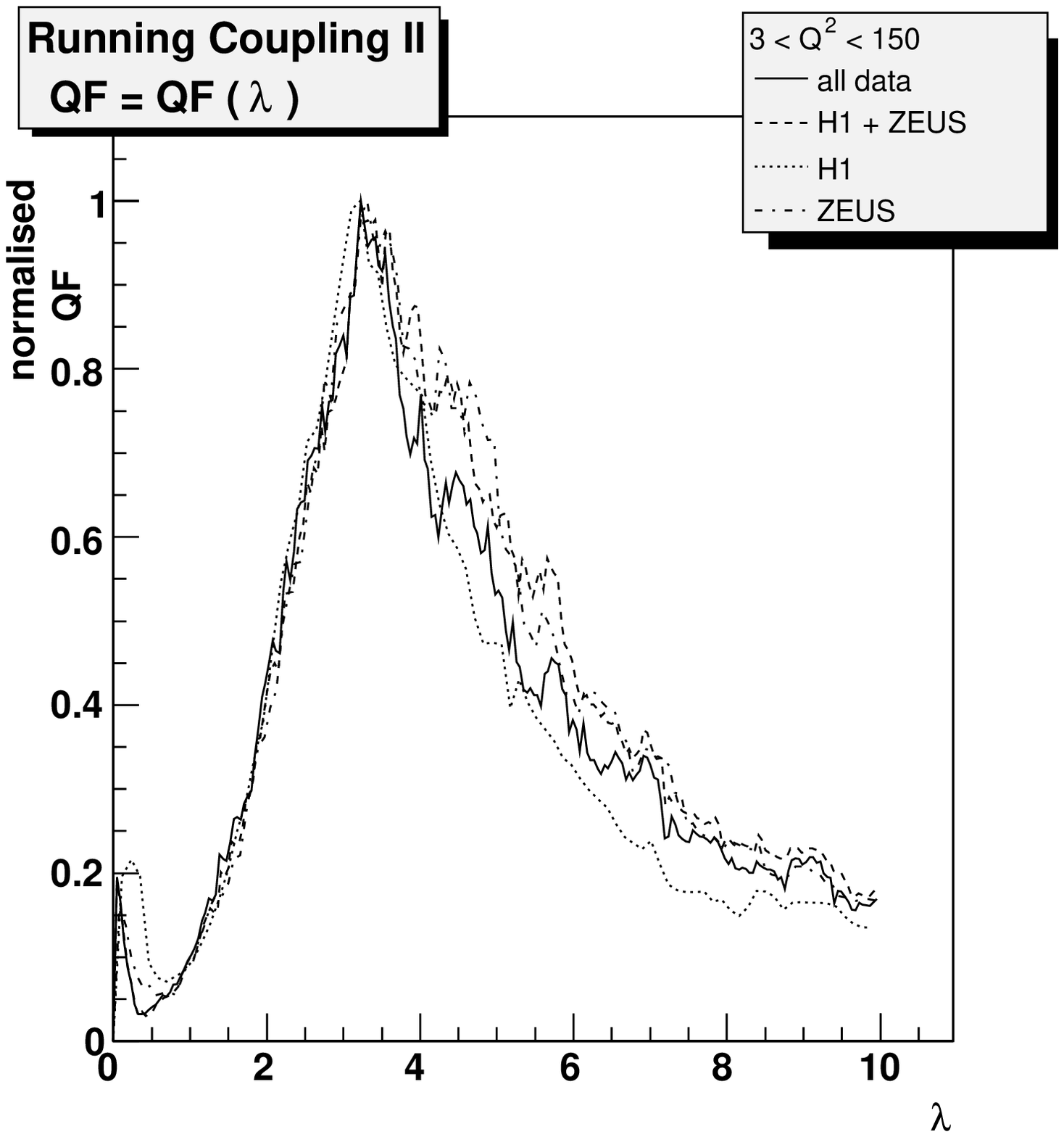} & \includegraphics[width=9cm]{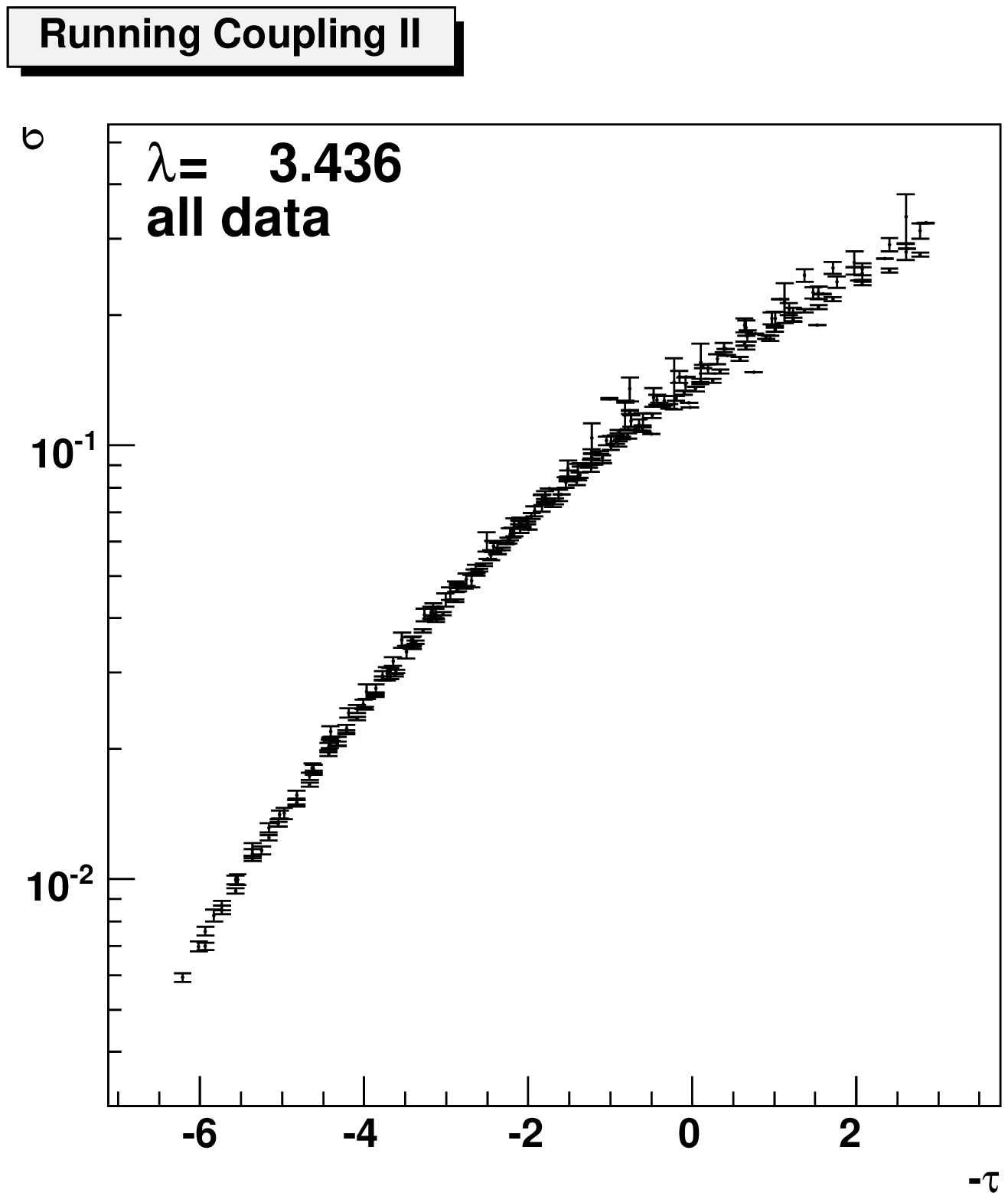}\\
\end{tabular}
\caption{\label{fig:pheno} Phenomenological results for the new running coupling scaling variable $\tau_n$ \eqref{RC2pheno}, with $Y_0=0$ and $\Lambda_{QCD}=0.2$ GeV. \emph{Left}: Normalized quality factor as a function of the parameter $\lambda$ for different data sets. \emph{Right}: Inclusive DIS cross section as a function of the scaling variable $\tau_n(Q^2,Y)$.}
\end{center}
\end{figure}
The results for the new variable are displayed in FIG.\ref{fig:pheno}. On the right-hand plot, the data shows a nice scaling property with $\tau_n$, which is effectively as good as  with $\tau_f$ \cite{Stasto:2000er,Gelis:2006bs} or as with $\tau_g$ \cite{Gelis:2006bs}. On the left-hand plot, one remarks a sharp peak in the quality factor, showing that the parameter $\lambda$ is determined unambiguously. That peak is remarquably stable when the study is done on some subsets of the data, like H1+ZEUS data only, or H1 only, or ZEUS only. It illustrates the robustness of the quality factor fit in that case.

When introducing the other parameters $Y_0$ and $\Lambda$ in the quality factor fit, we find that for the running coupling geometric scaling variable $\tau_g$, the optimal value for $Y_0$ remains compatible with zero. By contrast, for the new scaling with $\tau_n$, letting $\Lambda$ and $Y_0$ free improves significantly the quality factor from $QF=1.69$ to $QF=1.82$ (in $10^{-3}$ units). The optimal values of the additional parameters are $Y_0=-1.2$ and $\Lambda=0.3$ GeV, which are reasonably moderate shits. Hence, the prefered scaling behavior found in the inclusive DIS cross section data is obtained using the new scaling variable $\tau_n$. This new scaling works as well for DVCS, vector meson production, and diffractive data \cite{BPRS}.



\section{Conclusion and outlook}

We have shown that the Balitsky-Kovchegov equation with running coupling has no \emph{exact} scaling solution, contrary to the fixed coupling case. However, that equation is compatible with two types of approximate scaling solutions, which we have analyzed. The first type of scaling behavior boils down to the running coupling geometric scaling \eqref{SatScaleOldSc}, already known \cite{Iancu:2002tr}, while the second one \eqref{SatScaleNewSc} is new, and goes beyond the notion of geometric scaling.  

Using the traveling wave method, we have derived the asymptotic traveling wave solution (\ref{solLEnewscI},\ref{SatScaleNewScI}) associated with the new scaling variable \eqref{SatScaleNewSc}. Both running coupling traveling wave solutions are shown to be compatible concerning the evolution of the saturation scale, and the shape of the leading edge of the pulled front, while they differ in the front interior. It means that the asymptotic traveling wave solutions associated with the two scaling variables are both approximations of a same exact solution. The asymptotic shape of the leading edge and the evolution of the saturation scale are universal features of the exact solution, and the traveling wave method allows to reach them, within both approximations, because both approximate scalings become exact asymptotically. 

By contrast, other features such as the shape of the front closer to the saturation scale are more difficult to calculate within the traveling wave framework. One finds that the front is steeper (see FIG.\ref{fig:RCshape}) for the new solution \eqref{solLEnewscI} than for the geometric scaling solution \eqref{solLEoldsc1}. As the two approximate solutions rely on similar but distinct approximations, one could estimate their uncertainty by their difference. The exact solution is presumably lying between the two approximate solutions. Hence, it should have a steeper front than the running coupling geometric scaling solution, and than the fixed coupling solution, as found in numerical simulations  \cite{Albacete:2004gw,Albacete:2007yr}.\\

In the fixed coupling (and mean field) case, the main geometric scaling violations are the universal ones resulting from the convergence towards the asymptotic behavior. In the running coupling case, the dipole target amplitude shows simultaneously the two scaling behaviors  \eqref{SatScaleOldSc} and \eqref{SatScaleNewSc}. As these two scalings are compatible only in a large $L$ and large $Y$ limit, that property should bring violations of both scalings at nonasymptotic energies, in addition the universal scaling violations.
Those stronger scaling violations at running coupling, and the fact that the new scaling is slightly favoured by the DIS data \cite{BPRS}, explain why the running coupling geometric scaling is not significantly better than the fixed coupling geometric scaling at HERA energies.

The NLL and higher order terms are quantitatively important at HERA energies \cite{Triantafyllopoulos:2002nz}. The phenomenological determination of the scaling parameters done in \cite{BPRS} and in section \ref{sec:Pheno} may be useful to understand the nonasymptotic effects of those contributions.\\

Let us briefly discuss our various theoretical approximations. Following \cite{Dumitru:2007ew}, we have neglected Pomeron loops effects, which seem irrelevant in dense-dilute collisions in the running coupling case. Our result (\ref{solLEnewscI},\ref{SatScaleNewScI}) is free from NLL corrections. They appear indeed at the next order in the leading edge expansion \eqref{solLEnewscI} only. 
However, NLL and higher orders are phenomenologically important, as mentionned previously.
Our choice for the running coupling scale differs from the best ones \cite{Balitsky:2006wa,Kovchegov:2006vj}, but this difference is only marginally relevant in numerical simulations \cite{Albacete:2007yr}.
The dipole or parton correlations present in the Balitsky-JIMWLK equations, and the off-diagonal momentum transfert dependance,  do not appear in the linear part of the equation. Hence, the universality of the pulled front solutions guaranties that these correlations cannot modify our result (\ref{solLEnewscI},\ref{SatScaleNewScI}).  Note that the effect of all those approximations is obviously the same for the previous approximate solution (\ref{solLEoldsc1},\ref{SatScaleOldSc1}).\\

As an outlook, it would be interesting, on the theoretical side, to explore the tail of the front. The two scaling behaviors are no more relevant when $s_n$ \eqref{SatScaleNewSc} or $s_g$ \eqref{SatScaleOldSc} are of order $L$, \emph{i.e.} in the forward region of the front, where the initial condition still propagates. Hence, the crossover between the pulled front and the forward region seems more complex than in the fixed coupling case. It should be also important to calculate the next order of the leading edge expansion of the two running coupling solutions, including NLL effects. Moreover, the variable $s_n$ may be useful to study the solutions of the (LL or NLL) BFKL equation with running coupling.

On the phenomenological side, a systematic study \cite{BPRS}  of scaling properties of inclusive and exclusive DIS cross sections has been conducted simultaneously to the present paper. A further step would be to use our results to build a running coupling saturation model, in a similar way as the Iancu-Itakura-Munier model \cite{Iancu:2003ge} for the fixed coupling case. Most available phenomenological models of saturation are built at fixed coupling. Although successful to describe HERA data, they may fail to give reasonable predictions for the LHC, due to the different evolution of the saturation scale. This justifies the need for a running coupling saturation model, \emph{e.g.} based on the new scaling variable.

\begin{acknowledgments}
I would like to thank Robi Peschanski for useful remarks and the careful reading of the manuscript. I thank him, Christophe Royon and David \v S\'alek for a fruitful collaboration.

\end{acknowledgments}

\vspace{1cm}


\appendix

\section{Derivation of the  leading-edge of the new asymptotic solution}
\label{sec:appA}

We finish in this appendix the calculation of section \ref{sec:UnivAsymptScalingSol}, starting from Eq. \eqref{BKRCscLE3}

\be
0 = \chi'(\gc)  \left[\f{s_n}{L}\d_{s_n} -\d_L - \f{\gc s_n}{ L} \right]
f_{LE}(s_n,L)+\sum_{p=2}^\infty \f{1}{p!} \chi^{(p)}(\gc)
\left[-\left(1-\f{s_n}{L}\right)\d_{s_n} -\d_L - \f{\gc s_n}{ L} \right]^p f_{LE}(s_n,L)
\, .\label{BKRCscLE3A}
\ee

As discussed in section \ref{sec:UnivAsymptScalingSol}, one can factorize $f_{LE}$ as
\be
f_{LE}(s_n,L)  \simeq e^{-3\gc \beta L^{1/3}} G(z) \, , \label{ansatzLE}
\ee
$z$ being
\be
z \equiv \f{\bar{s}_n}{L^{\f{1}{3}}}  =\f{s_n}{L^{\f{1}{3}}} + 3 \beta  \label{scalingZ}
\, ,
\ee
with the shifted scaling variable
\be
\bar{s}_n \equiv s_n + 3 \beta L^{\f{1}{3}} =\f{L}{2}-\f{v_c (Y-Y_0)}{2 b L} + 3 \beta
L^{\f{1}{3}} \nonumber \, .
\ee
Inserting \eqref{ansatzLE} in the equation \eqref{BKRCscLE3A} written in $z$ and
$L$ variables, \emph{i.e.} doing the replacements
\ba
\d_{s_n} &\rightarrow& L^{-\f{1}{3}} \ \d_z \nn
\d_L &\rightarrow& \d_L - \f{z-3 \beta}{3 L} \ \d_z \, , \nonumber
\ea
one gets
\be
\chi'(\gc)  \left[-\f{4(z-3 \beta)}{3L}\d_z + \f{\gc (z-4 \beta)}{L^{\f{2}{3}}}
\right]G(z) \simeq \sum_{p=2}^\infty \f{(-1)^p}{p!} \chi^{(p)}(\gc)
\left[L^{-\f{1}{3}} \d_z -\f{4(z-3 \beta)}{3L}\d_z + \f{\gc (z-4
\beta)}{L^{\f{2}{3}}}  \right]^p G(z) \, .\label{BKRCscLE5}
\ee
$G(z)$ is determined by the terms of order $L^{-2/3}$, for $z={\cal O}(1)$.
Hence we have
\be
\chi'(\gc) \ \gc (z-4 \beta)\ G(z) =  \f{1}{2}\ \chi''(\gc)\ G''(z) \,
.\label{BKRCscLE6}
\ee

The following terms in \eqref{BKRCscLE5} are of order $L^{-1}$. The subleading
term in $f_{LE}$ is thus suppressed by a factor $L^{-1/3}$, compared to $G(z)$.
A factor $L^\delta$ appears when calculating the subleading term in $f_{LE}$, in
the same way as the factor $\exp(-3\gc \beta L^{1/3})$ appears when calculating
the leading term.
Our choice is to single out an $L^{1/3}$ factor from this $L^\delta$, and to
write the remaining part as a ${\cal O}(\log L)$ correction to the scaling
variable. Finally, $f_{LE}$ writes
\be
f_{LE}(s,L)  = e^{-3\gc \beta L^{1/3} + {\cal O}(\log L)} \ L^{1/3}  \left[G(z)
+ L^{-\f{1}{3}} H(z) + {\cal O}\left(L^{-\f{2}{3}}\right) \right] \, ,
\label{ansatzLEexact}
\ee
for $L$ large, and $z={\cal O}(1)$. The function $H(z)$ would be determined at
the next order, as the ${\cal O}(\log L)$ correction to the scaling variable.

Let us now calculate $G(z)$ and $\beta$ from the equation \eqref{BKRCscLE6}. Up
to an affine change of the variable $z$, this equation can be mapped into the
Airy equation. Hence $G(z)$ writes
\be
G(z)=C_1 \ \textrm{Ai}\left(\f{z-4 \beta}{D^{\f{1}{3}}} \right)+C_2 \
\textrm{Bi}\left(\f{z-4 \beta}{D^{\f{1}{3}}} \right) \, , \label{genericG}
\ee
where $\textrm{Ai}$ and $\textrm{Bi}$ are the Airy functions, $C_1$ and $C_2$
are two integration constants and $D$, playing the role of a diffusion
coefficient, is defined by
\be
D\equiv \f{\chi''(\gc)}{2\ \gc \chi'(\gc)}= \f{\chi''(\gc)}{2\ \chi(\gc)} \, .
\label{coeffD}
\ee
Our calculations are valid only if $s_n\ll \sqrt{L}$. \emph{A priori}, it allows
$\bar{s}_n$ to be large compared to $(DL)^{1/3}$, and then $z$ to be arbitrary large
for $L\rightarrow \infty$. Thus, $G(z)$ has to be regular for $z \rightarrow
\infty$, which gives $C_2 =0$.
The other boundary of the interval of validity of the solution
\eqref{ansatzLEexact} is located at $z\simeq 0$, and corresponds to the
transition to the saturation regime. For $1\ll \bar{s}_n\ll L^{1/3}$, the solutions
\eqref{ansatzLEexact} and \eqref{UnivFrontIntSol} are both valid, and then we
have to match them. We have determined that the scaling is modified by the
leading edge dynamics from $s_n$ to $\bar{s}_n$. And in the front interior, the
difference between $s_n$ and $\bar{s}_n$ is a subleading effect for the solution
\eqref{UnivFrontIntSol}. Hence, we can match the leading edge solution
\be
\bar{N}(s_n,L)  = e^{-\gc \bar{s}_n + {\cal O}(\log L)} \ L^{\f{1}{3}}  \left[C_1 \
\textrm{Ai}\left(\f{\bar{s}_n}{(DL)^{\f{1}{3}}}  - \f{4 \beta}{(D)^{\f{1}{3}}}
\right) + L^{-\f{1}{3}} H\left(\f{s_1}{L^{\f{1}{3}}}\right) + \dots \right] \,
\label{solLE1}
\ee
with the front interior solution with $s_n$ replaced by $\bar{s}_n$
\be
\bar{N}_0(\bar{s}_n)= A\ (\bar{s}_n+B) \ e^{-\gc \bar{s}_n}\, . \label{solFI1}
\ee
For $1\ll \bar{s}_n\ll L^{1/3}$, \eqref{solLE1} can be expanded as
\be
\bar{N}(s_n,L)  = e^{-\gc \bar{s}_n + {\cal O}(\log L)} \ L^{\f{1}{3}}  \left[C_1 \
\textrm{Ai}\left(- \f{4 \beta}{D^{\f{1}{3}}}  \right)
+C_1 \ \f{\bar{s}_n}{(DL)^{\f{1}{3}}} \ \textrm{Ai}'\left(- \f{4
\beta}{(D)^{\f{1}{3}}}\right)
+ L^{-\f{1}{3}} H\left(0\right) + \dots \right] \, .  \label{solLE2}
\ee
Hence, one finds that $-4 \beta /D^{1/3}$ has to be a zero of the Airy function
Ai. Requiring that $\bar{N}(s_n,L)$ is positive for $z>0$ selects the rightmost
zero $\xi_1\simeq -2.338$. Hence, one has
\be
\beta=-\f{1}{4}\ \xi_1\ D^{\f{1}{3}} \, .
\ee
The matching of the terms proportional to $\bar{s}_n$ gives
\be
C_1=\f{D^{1/3}}{\textrm{Ai}'(\xi_1)} \ A \, .
\ee
The constant terms give then $H(0) = A B$. However, one can replace $\bar{s}_n$ by
$\bar{s}_n + B$ in the argument of the Airy function Ai in \eqref{solLE1}. It
corresponds to a choice of the repartition of the higher order terms, very
similar to the choice of the $L^{1/3}$ in front of the solution \eqref{solLE1}.
Using this prescription, one has $H(0)=0$, and then the contribution of the
higher orders is reduced, at least in the neighborhood of $z=0$.

Finally, the universal asymptotic traveling wave solution of the BK equation
with running coupling associated with the scaling \eqref{SatScaleNewScPar} writes, for large $L$ and $1\ll \bar{s}_n \ll \sqrt{L}$

\ba
N(L,Y)&=&\f{A}{\textrm{Ai}'(\xi_1)} (D L)^{1/3} e^{- \gc \bar{s}_n  +{\cal O}(\log
L)} \ \left[\textrm{Ai}\left(\xi_1+ \f{\bar{s}_n+B}{(D L)^{1/3}}\right)  +   L^{-1/3}
H\left(\f{\bar{s}_n}{L^{1/3}}\right)+\dots \right]\label{solLEnewscA} \\
\textrm{with} \quad \bar{s}_n&=&\f{L}{2}-\f{v_c (Y-Y_0)}{2 b L}- \f{3 \xi_1}{4} (D
L)^{1/3}\label{SatScaleNewScA}\, .
\ea

Note that we have not determined the parameters $A$ and $B$. However, they are
not completely free, and they should be constrained by matching the expression
\eqref{UnivFrontIntSol} with numerical, or analytical if possible, solution of
the nonlinear equation \eqref{BKRCFI0} with $v=v_c$. Note also that the prefactor $(D L)^{1/3}$ is arbitrary, as it is part of the still unknown ${\cal O}(\log
L)$ term in the exponent. We have written it just for convenience, in order to cancel in the large $L$ limit the $(D L)^{1/3}$ denominator present in the Airy function. The same is true for the $Y^{1/6}$, or $t^{1/3}$,  prefactor in the solution associated the variable $s_g$ in \cite{Mueller:2002zm,Munier:2003sj}.

\bibliography{MaBiblio}

\end{document}